\documentclass[12pt]{article}
\usepackage{graphicx}
\begin{document}
\begin{center}

{\Large \bf Photons in the Quantum World}

\vspace{10mm}

Sibel Ba{\c s}kal  \\
Department of Physics, Middle East Technical University, 06800 Ankara, Turkey \\[5ex]

Young S. Kim \\
Center for Fundamental Physics, University of Maryland College Park,\\ Maryland,
MD 20742, USA \\[5ex]

Marilyn E. Noz \\
Department of Radiology, New York University, New York, NY 10016, USA

\end{center}

\vspace{10mm}

\abstract{
Einstein's photo-electric effect allows us to regard
electromagnetic waves as massless particles.  Then, how is the
photon helicity translated into the electric and magnetic fields
perpendicular to the direction of propagation?  This is an issue of
the internal space-time symmetries defined by Wigner's little group
for massless particles.  It is noted that there are three generators
for the rotation group defining the spin of a particle at rest.  The
closed set of commutation relations is a direct consequence of
Heisenberg's uncertainty relations.  The rotation group can be
generated by three two-by-two Pauli matrices
for spin-half particles.  This group of two-by-two matrices is called
$SU(2)$, with two-component spinors.  The direct product of two spinors
leads to four states leading to one spin-0 state and one spin-1 state
with three sub-states.  The $SU(2)$ group can be expanded to another
group of two-by-two matrices called $SL(2,c)$, which serves as the
covering group for the group of Lorentz transformations.  In this
Lorentz-covariant regime, it is possible to Lorentz-boost the particle
at rest to its infinite-momentum or massless state.  Also in this
$SL(2,c)$ regime, there are four spin states for each particle, as in
the case of the Dirac equation.  The direct product of two $SL(2,c)$
spinors thus leads sixteen states.  Among them, four of them can be used
for the electromagnetic four-potential, and six for the Maxwell tensor.
The gauge degree of freedom is shown to be a Lorentz-boosted rotation.
The polarization of massless neutrinos is interpreted as a consequence
of the gauge invariance.
}

\vspace{30mm}
Invited paper presented at the 15th International Conference on Squeezed States and
Uncertainty Relations (Jeju, South Korea, August 2017)

\newpage

\section{Introduction}

The algebra of quantum mechanics starts from Heisenberg's  commutation relations
\begin{equation} \label{heisen01}
\left[x_{i}, p_{j}\right] = i \delta_{ij},
\end{equation}
with
\begin{equation}
  p_{i} = -i\frac{\partial}{\partial x_{i}} .
\end{equation}
These expressions are well known.

Let us next consider the operators
\begin{equation}\label{rr00}
J_{i} = -i \epsilon_{ijk}  x_{j} \frac{\partial}{\partial x_{k}} =
-i \left(x_{j}\frac{\partial}{\partial x_{k}} - x_{k}\frac{\partial}{\partial x_{j}} \right),
\end{equation}
which satisfy the commutation relations
\begin{equation}\label{rr11}
  \left[J_{i}, J_{j}\right] = i\epsilon_{ijk} J_{k} .
\end{equation}
This closed set of commutation relations is the Lie algebra of the three-dimensional
rotation group $O(3)$.  This Lie algebra is a direct consequence of Heisenberg's
uncertainty relations given in Eq.(\ref{heisen01}).
\par
The simplest matrices representing this Lie algebra are
\begin{equation}\label{spin11}
S_{i} = \frac{1}{2}~\sigma_{i} ,
\end{equation}
where $\sigma_{i}$ are the two-by-two Pauli spin matrices.  We use the spinors
\begin{equation}
  u = \pmatrix{1 \cr 0}, \quad\mbox{and}\quad
                                        v = \pmatrix{0 \cr 1} ,
\end{equation}
for the spin-up and spin-down states, respectively.

\par
With these spinors, we can construct the spin-0 and spin-1 states in the following
manner.  For the spin-0 state, we make the anti-symmetric combination
\begin{equation}
   \frac{1}{\sqrt{2}}(uv - vu).
\end{equation}
There are three spin-1 states.  They are
\begin{equation} \label{spin33}
 uu, \quad   \frac{1}{\sqrt{2}}(uv + vu), \quad vv,
\end{equation}
for the $z$-component spin 1, 0, and -1 respectively.

Next, we all know photons are massless particles with spin 1 parallel or anti-parallel
to its momentum. They are called helicities.  We are also familiar with the expressions
\begin{equation} \label{max01}
A_{\mu} = \pmatrix{A_{0} \cr A_{z} \cr A_{x} \cr A_{y}},
    \quad\mbox{and}\quad
F_{\mu\nu} = \pmatrix{0 & -E_{z} & -E_{x} & -E_{y} \cr  E_{z} & 0 & -B_{y}  & B_{x} \cr
                      E_{x} & B_{y} & 0 & -B_{z} \cr E_{y} & -B_{x} & B_{z} & 0 }.
\end{equation}
These are Maxwell's four-vector and second-rank four-tensor for electromagnetic fields.
These expressions belong to Einstein's Lorentz-covariant world.  For convenience, we use
the Minkowskian four-vector convention of $(t, z, x, y)$ throughout the paper.

Here is the question.  Is it possible to derive these Maxwell vector and tensor from
Heisenberg's relations given in Eq.(\ref{heisen01})?  The answer is No, but is it
possible to construct a bridge between them?   The answer is Yes, but this question
has a stormy history.  The purpose of this paper is to provide a simple answer to
this question.   The bridge consists of the set of three two-by-two ``imaginary''
Pauli matrices $i~\sigma_{i}$:
\begin{equation}\label{boost11}
K_{i} = \frac{i}{2}~\sigma_{i} .
\end{equation}
They correspond to
\begin{equation}
K_{i} = -i\left(t\frac{\partial}{\partial x_{i}} + x_{i}\frac{\partial}{\partial t}\right) ,
\end{equation}
applicable to the four-dimensional Minkowskian space.

The six matrices consisting of $S_{i}$ and $K_{i}$ become the generators of the group
$SL(2,c)$ isomorphic to the group of Lorentz transformations.  This group thus allows
us to Lorentz-transform spinors which will eventually lead us to the electromagnetic
four-vector and four-tensor.

It was of course Einstein who unified the energy-momentum relation for both massive
and massless particles.  We are now faced with the problem of unifying internal space-time
symmetries.  Einstein's photon has spin-one parallel or anti-parallel to its momentum.
For a particle at rest, we all know how to construct spin-1 states from two spinors as
was the case in Eq.(\ref{spin33}).  The issue is how to Lorentz-boost those spin-1 states
to reach the Maxwell tensor and four-vector for electromagnetic field.

\par
It was Eugene Wigner who pioneered this research line.  In 1939~\cite{wig39}, he
constructed the subgroups of the Lorentz group whose transformations leave the
four-momentum of a given particle invariant.  He called these subgroups ``little groups.''
Thus, Wigner's little groups dictate the internal space-time symmetries of particles in
the Einstein's Lorentz-covariant world which includes both massive and massless particles,
as shown in Table~\ref{emc}.  This table was first published in 1986~\cite{hks86jmp}.
Indeed, the photon polarization and the gauge degree of freedom are the issues of
the internal space-time symmetries of massless particles.

\begin{center}
\begin{table}
\caption{Extension of the concepts contained in Einstein's $E = mc^{2}$ to photons helicity
and gauge.  Under the Lorentz boost along the $z$ direction, the $z$ component of the spin
remains as the helicity, but the transverse components collapse into one gauge degree of
freedom.}\label{emc}
\vspace{2mm}
\begin{center}
\begin{tabular}{cccccccc}
\hline\\ [-2.3ex]
\hline\\ [-1.2ex]
 \hspace{3mm}  &\hspace{5mm} & Slow & \hspace{5mm}& Relativistic &\hspace{3mm}
 & Fast & \hspace{2mm}\\[1.0ex]
\hline\\[0.6ex]
\hspace{2mm} $\matrix{\mbox{Energy-}\cr \mbox{momentum}}$  &\hspace{3mm}&
$ p^{2}/2m  $  &\hspace{3mm}&  $\sqrt{p^{2} + m^{2}}$ &\hspace{3mm}& $E = p$
&\hspace{2mm}
\\ [3.0ex]
\hline \\[-0.3ex]
\hspace{2mm} $\matrix{\mbox{Helicity} \cr \mbox{ Spin \& Gauge}}$  & \hspace{3mm} &
$\matrix{S_{3} \cr S_{1}, S_{2}} $  & \hspace{3mm} &
$\matrix{\mbox{Wigner's} \cr \mbox{Little Groups}}$
& \hspace{3mm}
 & $\matrix{\mbox{Helicity} \cr \mbox{Gauge Trans.}}$ & \hspace{2mm}
 \\[3.0ex]
\hline \\ [-2.3ex]
\hline
\end{tabular}
\end{center}
\end{table}
\end{center}

In Sec.~\ref{lgroup}, we present two-by-two representation of the Lorentz group, and
Wigner's little groups in Sec.~\ref{wigner}. In Sec.~\ref{massless}, we discuss
massless particles as large-momentum or small-mass limit of massive particles.
It is shown that there are four spin states in the Lorentz-covariant world.  Thus
there are sixteen different ways to combine two spinors.  In Sec.~\ref{svt}, we
construct explicitly those sixteen states.  Among them are the electromagnetic
four-vector and the Maxwell tensor.  It is pointed out that the polarization of massless
neutrinos is a consequence of gauge invariance.

\section{Lorentz Group and Its Representations}\label{lgroup}

In addition to the rotation generators of Eq.(\ref{rr00}), we can consider another set
of three operators, namely
\begin{equation}
K_{i} = -i \left(x_{i}\frac{\partial}{\partial t} + t \frac{\partial}{\partial x_{i}} \right) .
\end{equation}
These operators are known to generate Lorentz boosts in the Minkowskian space of one time
direction and three spatial dimensions, and they satisfy the commutation relations
\begin{equation}\label{bb11}
  \left[K_{i}, K_{j}\right] = -i\epsilon_{ijk} J_{k} .
\end{equation}
These three boost generators do not lead to a closed set of commutation relations.  However,
with the $J_{i}$ generators, they satisfy the commutation relations
\begin{equation} \label{rb11}
  \left[J_{i}, K_{j}\right] = i\epsilon _{ijk} K_{j} .
\end{equation}

Let us write the commutation relations of Eqs.(\ref{rr11}), (\ref{bb11}) and (\ref{rb11})
as one closed set of commutation relations
\begin{equation}\label{lg11}
  \left[J_{i}, J_{j}\right] = i\epsilon_{ijk} J_{k}, \quad
  \left[J_{i}, K_{j}\right] = i\epsilon_{ijk} K_{k} , \quad
  \left[K_{i}, K_{j}\right] = -i\epsilon_{ijk} K_{k} .
\end{equation}
This set is called the Lie algebra of the Lorentz group.

In terms of four-by-four matrices these generators take the form:
\begin{equation}\label{gene01m}
J_{3} = \pmatrix{0 & 0 & 0 & 0 \cr 0 & 0 & 0 & 0 \cr
0 & 0 & 0 & -i \cr 0 & 0 & i & 0 }, \qquad
K_{3} = \pmatrix{ 0 & i & 0 & 0 \cr i & 0 & 0 & 0 \cr
0 & 0 & 0 & 0 \cr 0 & 0 & 0 & 0 },
\end{equation}
for rotations around and boosts along the $z$ direction, respectively.
Here, again the ordering of the coordinates in Minkowskian space-time
are $(t, z, x, y)$, where the transformation matrices corresponding
to the generators above are applicable.

Similar expressions can be written for the $x$ and $y$ directions.
We see that the rotation generators $J_{i}$ are Hermitian, but
the boost generators $K_{i}$ are anti-Hermitian.
\par

Four-by-four matrices that leave the quantity $\left(t^2 - z^2 - x^2 - y^2 \right)$
invariant, in the four-dimensional Minkowski space forms the basis of the group
of Lorentz transformations.
Since there are three rotation and three boost generators, the Lorentz group is a
six-parameter group.

The Lorentz group can also be represented by two-by-two matrices.  If we choose
\begin{equation}\label{lg22}
 J_{i} = \frac{1}{2}\sigma_{i}, \quad
 K_{i} = \frac{i}{2}\sigma_{i}.
 \end{equation}
They satisfy the set of commutation relations given in Eq.(\ref{lg11}). Thus, to each
two-by-two transformation matrix, there is a corresponding four-by-four matrix
applicable to the Minkowskian space.

The algebra of Eq.(\ref{lg22}) is invariant under the sign change of the $K_{i}$ matrices.
Let us introduce the notation
\begin{equation}
 \dot{K_{i}} = - K_{i} .
\end{equation}
Then
\begin{equation}\label{lg22dot}
 J_{i} = \frac{1}{2}\sigma_{i}, \qquad \dot{K}_{i} = -\frac{i}{2}\sigma_{i} .
\end{equation}
Corresponding to these two-by-two matrices, we can construct one set of two-component
spinor (spin-up and spin-down) for the undotted representation, and another set for
the dotted representation. There are thus four spin states in the Lorentz-covariant
world as shown in Table~\ref{spinors}.  This is the reason why the Dirac spinor has
four components~\cite{knp86}.

\begin{center}
\begin{table}
\caption{Spinors in the relativistic world.  The spinors $u$ and $v$ are for spin-up and
spin-down states respectively.  Under the Lorentz boost, the dotted spinors are boosted in
the opposite direction.}\label{spinors}
\vspace{2mm}
\begin{center}
\begin{tabular}{cccccc}
\hline\\ [-2.3ex]
\hline\\ [-1.2ex]
 \hspace{3mm}  &\hspace{5mm} & undotted & \hspace{5mm}& dotted & \hspace{2mm}\\[1.0ex]
\hline\\[0.6ex]
\hspace{3mm} Spin up  &\hspace{5mm}&
       $    u  $ &\hspace{5mm}&  $\dot{u}$ & \hspace{2mm}
\\ [2.0ex]
\hline \\[-0.3ex]
\hspace{3mm} Spin down  &\hspace{5mm}&
$ v $
&\hspace{5mm} &
$\dot{v}$& \hspace{2mm}
\\[2.0ex]
\hline \\ [-2.3ex]
\hline
\end{tabular}
\end{center}
\end{table}
\end{center}

As far as rotations are concerned, the representation constructed from the Lie
algebra of Eq.(\ref{lg22dot}) is transformed in the same way as that of
Eq.(\ref{lg22}).  However, the Lorentz boosts are performed in opposite
directions.

\par
If two spinors are coupled, there are 16 (= 4 x 4) states, which can be partitioned
into to the spin-0 and spin-1 states.  We shall come back to this problem in
Sec.~\ref{svt}.

\section{Wigner's Little Groups}\label{wigner}

In 1939~\cite{wig39},
Wigner~considered the subgroups of the Lorentz group the transformations
of which leave the four-momentum of a given particle invariant.
It is well-known that the momentum of a massive particle at rest
is invariant under rotations.  On the other hand massless particles do not
have rest frames. Then how shall we study the
space-time symmetries of massless particles?  In this paper we shall address this
problem, primarily focusing on the case of the photons.

\par
The generators of Eq.(\ref{lg11}) lead to the group of two-by-two unimodular
matrices of the form
\begin{equation}\label{alphabeta}
 G = \pmatrix{\alpha & \beta \cr \gamma & \delta} ,
\end{equation}
whose elements are complex numbers, with $\det(G) = 1$.
Six independent real parameters are required to generate a group,
where its six generators are given in Eq.(\ref{lg22}).
This type of matrices form a group and are called $SL(2,c)$.

The generators $K_{i}$ are not Hermitian, therefore the matrix $G$ is not always
unitary.  Moreover, its Hermitian conjugate is not necessarily its inverse.
This two-by-two representation has extensively been studied in the
literature~\cite{dir45dub,barg47,naimark54,dlug09,bkn15,bkn17symm}.

\par
In this two-by-two representation, the space-time four-vector can be written as
\begin{equation}\label{slc01}
\pmatrix{t + z & x - iy \cr x + iy & t - z},
\end{equation}
whose determinant is $t^{2} - z^{2} - x^{2} - z^{2}$, and remains invariant under
the Hermitian transformation:
\begin{equation}\label{slc02}
X' = G~X~G^{\dag} .
\end{equation}
Therefore this is a Lorentz transformation, which can explicitly
be written as:
\begin{equation}\label{slc04}
\pmatrix{t' + z' & x' - iy' \cr x' + iy' & t' - z'} =
\pmatrix{\alpha & \beta \cr \gamma & \delta}
\pmatrix{t + z & x - iy \cr x + iy & t - z}
\pmatrix{\alpha^* & \gamma^* \cr \beta^* & \delta^*} .
\end{equation}

The transformation matrices of the Lorentz group applicable to four
dimensional Minkowskian space-time can be constructed with these six
independent real parameters of $Sl(2,c)$~\cite{knp86,bkn15}.
In the sequel we shall only be using the two-by-two representations
of the Lorentz group.  We give them in Table~\ref{tab11}.

\begin{table}
\caption{Two-by-two representations of the Lorentz group.  Rotations take the same form
for both dotted and undotted representations, but  boosts are performed in opposite
directions. } \label{tab11}
\vspace{0.5ex}
\begin{center}
\begin{tabular}{lcllc}
\hline \\ [-2.3ex]
\hline\\[-0.4ex]
\hspace{3mm}Generators &\hspace{1mm} & $ \matrix{\mbox{Transformation Matrices for}\cr
                                 \mbox{Undotted Representation} }$ & \hspace{2mm} &
 $\matrix{\mbox{Transformation Matrices for} \cr
                  \mbox{Dotted Representation} }$ \\[2.0ex]
\hline\\ [0.8ex]
 $J_{3} = \frac{1}{2}\pmatrix{1 & 0 \cr 0  & -1}$ &\hspace{2mm} &
  $\pmatrix{\exp{(-i\phi/2)} &  0    \cr 0 & \exp{(i\phi/2)}} $  & same &
  $\pmatrix{\exp{(-i\phi/2)} &  0    \cr 0 & \exp{(i\phi/2)}} $
\\[4.0ex]
 $K_{3} = \frac{1}{2}\pmatrix{i & 0 \cr 0 & -i}$ &\hspace{2mm} &
   $\pmatrix{\exp{(\eta/2)} &  0 \cr 0 & \exp{(-\eta/2)}}$     & inverse &
    $\pmatrix{\exp{(-\eta/2)} &  0 \cr 0 & \exp{(\eta/2)}}$
\\[3.0ex]
\hline\\ [-0.5ex]
$J_{1} = \frac{1}{2}\pmatrix{0 & 1 \cr 1 & 0}$  &  \hspace{2mm} &
   $\pmatrix{\cos(\theta/2) &  i\sin(\theta/2) \cr i\sin(\theta/2) & \cos(\theta/2)}$ & same &
   $\pmatrix{\cos(\theta/2) &  i\sin(\theta/2) \cr i\sin(\theta/2) & \cos(\theta/2)}$
\\[4.0ex]
$K_{1} = \frac{1}{2}\pmatrix{0 & i \cr i & 0}$ &\hspace{2mm} &
$\pmatrix{\cosh(\lambda/2) &  \sinh(\lambda/2)  \cr \sinh(\lambda/2) & \cosh(\lambda/2)}$ & inverse &
$\pmatrix{\cosh(\lambda/2) &  -\sinh(\lambda/2)  \cr -\sinh(\lambda/2) & \cosh(\lambda/2)}$
\\[3.0ex]
\hline\\ [-0.5ex]
$J_{2} = \frac{1}{2}\pmatrix{0 & -i \cr i & 0}$ & \hspace{2mm}&
  $\pmatrix{\cos(\theta/2) &  -\sin(\theta/2) \cr \sin(\theta/2) & \cos(\theta/2)}$ & same &
  $\pmatrix{\cos(\theta/2) &  -\sin(\theta/2) \cr \sin(\theta/2) & \cos(\theta/2}$
\\[4.0ex]
$K_{2} = \frac{1}{2}\pmatrix{0 & 1 \cr -1 & 0}$ &\hspace{2mm} &
 $\pmatrix{\cosh(\lambda/2) &  -i\sinh(\lambda/2) \cr i\sinh(\lambda/2) & \cosh(\lambda/2)}$ & inverse &
  $\pmatrix{\cosh(\lambda/2) &  i\sinh(\lambda/2) \cr -i\sinh(\lambda/2) & \cosh(\lambda/2)}$
\\[3.0ex]
\hline  \\ [-2.3ex]
\hline\\ [-0.5ex]
\end{tabular}
\end{center}
\end{table}

In a similar manner to that of Eq.(\ref{slc01}), the four-momentum can be expressed as
a two-by-two matrix in the form
\begin{equation}\label{mom22}
P = \pmatrix{p_0 + p_z & p_x - ip_y \cr p_x + ip_y & p_0 - p_z},
\end{equation}
where $p_0$ is defined through Einstein's equation
$p_{0}^{2}-(p_z^2 + p_x^2 + p_2^2)=m^2.$
The transformation property of Eq.(\ref{slc04}) is also applicable to this
energy-momentum matrix.  In 1939~\cite{wig39}, Wigner considered the following
two-by-two matrices:
\begin{equation}\label{slc06}
P_{+} = \pmatrix{1 & 0 \cr 0 & 1} , \qquad
P_{0} = \pmatrix{1 & 0 \cr 0 & 0} , \qquad
P_{-} = \pmatrix{1 & 0 \cr 0 & -1} .
\end{equation}
whose determinants are 1, 0, and $-$1, respectively, corresponding to the
four-momenta of massive, massless, and imaginary-mass particles, as shown
in Table~\ref{tab22}.
Wigner's little groups are the subgroups of the Lorentz
group whose transformations leave $P_{i}$ invariant:
\begin{equation}\label{wc00}
 W~P_{i}~W^{\dag} = P_{i} ,
\end{equation}
where $i = +, 0, -$.
Since the momentum of the particle is fixed, these little groups define the
internal space-time symmetries of the particle.

The rotation matrix around this axis is expressed as:
\begin{equation}\label{rotz}
 Z(\phi) = \pmatrix{e^{-i\phi/2} & 0 \cr 0 & e^{i\phi/2}}.
\end{equation}
Then we have
\begin{equation}
 Z(\phi)~ P_{i}~ Z^{\dag}(\phi) = P_{i} ,
\end{equation}
which means that for all the three cases the four-momentum remains invariant
under rotations around the $z$ axis.

\par
Let us now explicitly give the transformation matrices of the little groups for
i)~massive, ii)~massless and iii)~imaginary mass particles.

\par
{\bf Case i)}~~For a massive particle at rest whose momentum is $P_{+}$,
the little group is to satisfy:
\begin{equation}\label{wc01}
 W~P_{+}~W^{\dag} = P_{+} .
\end{equation}
This four-momentum remains invariant under rotations around the $y$ axis, whose
transformation matrix is
\begin{equation}\label{wm+}
  R(\theta) = \pmatrix{\cos(\theta/2) & - \sin(\theta/2) \cr
  \sin(\theta/2) & \cos(\theta/2)}.
\end{equation}
This matrix together with $Z(\phi)$ leads to the rotation also around the $x$ axis.  Thus,
Wigner's little group for the massive particle is the three-dimensional rotation subgroup
of the Lorentz group generated by $S_{i}$ given in Eq.(\ref{spin11}).

\par
{\bf Case ii)}~~ For a massless particle whose momentum is $P_{0}$
the triangular matrix of the form:
\begin{equation} \label{wm0}
T(\gamma) = \pmatrix{1 & -\gamma \cr 0 & 1} ,\quad\mbox{or} \quad
\dot{T}(\gamma) = \pmatrix{1 & 0 \cr \gamma  & 1}
\end{equation}
satisfies the Wigner condition of Eq.(\ref{wc00}).  For rotations
around the $z$ axis, these triangular matrices become
\begin{equation}\label{wm0z}
T\left(\gamma e^{-i\phi}\right) = \pmatrix{1 & -\gamma\exp{(-i\phi)} \cr 0 & 1}
\quad\mbox{or}\quad \dot{T}\left(\gamma e^{-i\phi}\right) = \pmatrix{1 & 0 \cr
\gamma\exp{(i\phi)}  & 1} . \end{equation}
The $T$ matrix is generated by:
\begin{equation}\label{n12}
N_{1} = J_{2} - K_{1} = \pmatrix{0 & -i \cr 0 & 0} ,
\quad\mbox{and}\qquad
N_{2} = J_{1} + K_{2} = \pmatrix{0 & 1 \cr 0 & 0} .
\end{equation}
Its dotted matrix is generated by
\begin{equation}\label{n12d}
\dot{N}_{1} = J_{2} + K_{1} = \pmatrix{0 & 0 \cr i &  0} ,
\quad\mbox{and}\qquad
\dot{N}_{2} = J_{1} - K_{2} = \pmatrix{0 & 0 \cr 1 & 0} .
\end{equation}
Thus, the little group is generated by $J_{3}$, $N_{1}$, and $N_{2}$. Together with $J_{3}$
they satisfy the following sets of commutation relations:
\begin{equation}\label{eucl11}
\left[N_{1}, N_{2} \right] = 0 , \qquad
\left[J_{3}, N_{1} \right] = i N_{2}, \qquad
\left[J_{3}, N_{2} \right] = -i N_{1} , \qquad
\end{equation}
and
\begin{equation}\label{eucl22}
\left[\dot{N}_{1}, \dot{N}_{2} \right] = 0 , \qquad
\left[J_{3}, \dot{N}_{1} \right] = i \dot{N}_{2}, \qquad
\left[J_{3}, \dot{N}_{2} \right] = -i \dot{N}_{1} . \qquad
\end{equation}

Wigner in 1939~\cite{wig39} observed that the first set given in Eq.(\ref{eucl11})
is the same as that of the generators for the two-dimensional Euclidean group with
one rotation and two translations. For massless particles rotation corresponds to the
helicity of the particle and it physical meaning is well understood.
On the other hand, the physical interpretation of $N_{1}$ and $N_{2}$ has not been
clarified, until it was completely resolved in 1990~\cite{kiwi90jmp}.
We now know that they generate gauge transformations~\cite{kuper76,hk81ajp,wein95}.

\par
{\bf Case iii)}~~For an imaginary mass particle whose momentum is $P_{-}$, the little
group matrix is of the form:
\begin{equation}\label{wm-}
S(\lambda) = \pmatrix{\cosh(\lambda/2) & \sinh(\lambda/2) \cr
 \sinh(\lambda/2) & \cosh(\lambda/2) },
\end{equation}
and satisfies the Wigner condition of Eq.(\ref{wc00}). This corresponds to
the Lorentz boost along the $x$ direction generated by $K_{1}$ as shown in
Table~\ref{tab11}. Because of the rotational symmetry around the $z$ axis,
condition in Eq.(\ref{wc00})is also satisfied by the boost along the $y$ axis.
Therefore, the little group is generated by $J_{3}, K_{1}$, and $K_{2}$,
and they satisfy the commutation relations:
\begin{equation}
 \left[J_{3}, K_{1} \right] = i K_{2}, \qquad
 \left[J_{3}, K_{2} \right] = -i K_{1}, \qquad
 \left[K_{1}, K_{2} \right] = -i J_{3} .
\end{equation}
This is the  $O(2,1)$ subgroup of the Lorentz group applicable to two
space-like and one time-like~dimensions. The same commutation relations
are satisfied for the dotted matrices.

\par
To summarize the setting, in Table~\ref{tab22}
we give Wigner momentum matrices along with their corresponding
transformation matrices.

\begin{table}
 \caption{Wigner four-momentum matrices.  Their two-by-two matrix forms are given,
  together with their corresponding transformation matrices in the dotted and undotted
  representations.  These momentum matrices have determinants that are positive, zero,
  and negative for the massive, massless, and imaginary-mass particles respectively
  respectively.}\label{tab22}
 \vspace{3mm}
 \begin{center}
 	\begin{tabular}{lllcccc}
 		\hline \hline\\[0.2ex]
 Mass & \hspace{2mm} & $\matrix{\mbox{ Four-} \cr \mbox{momentum}}$  &
   $ \matrix{\mbox{Transformation Matrices for}\cr \mbox{Undotted Representation} }$
   \hspace{1mm} & $\matrix{\mbox{Transformation Matrices for} \cr
   \mbox{Dotted Representation} }$ \\[2.0ex]
 \\[1.0ex]
 \hline\\[-0.5ex]
 Massive &\hspace{2mm} &
 $\pmatrix{1 & 0 \cr 0 & 1}$  &
 $\pmatrix{\cos(\theta/2) & -\sin(\theta/2)\cr  \sin(\theta/2) & \cos(\theta/2)}$ &
 $\pmatrix{\cos(\theta/2) & -\sin(\theta/2)\cr  \sin(\theta/2) & \cos(\theta/2)}$ \\[2.5ex]
       \hline\\[-1.0ex]
 Massless & \hspace{2mm} &
 $\pmatrix{1 & 0 \cr 0 & 0}$ &
 $\pmatrix{1 & -\gamma \cr 0 & 1}$   &
 $\pmatrix{1 & 0 \cr \gamma & 1}$ \\[2.5ex]
 \hline\\[-1.0ex]
 $\matrix{\mbox{Imaginary} \cr \mbox{mass} }$ & \hspace{2mm} &
 $\pmatrix{1 & 0\cr 0 & -1}$ &
 $\pmatrix{\cosh(\lambda/2) & \sinh(\lambda/2) \cr
 \sinh(\lambda/2) & \cosh(\lambda/2)}$ &
 $\pmatrix{\cosh(\lambda/2) & -\sinh(\lambda/2) \cr
 -\sinh(\lambda/2) & \cosh(\lambda/2)}$
 \\ [2.5ex]
 \hline
 \hline\\[-0.5ex]
 \end{tabular}
 \end{center}
\end{table}

\section{Massive and Massless Particles}\label{massless}

Indeed, the massive particle at rest remains invariant under rotations.  Let us
Lorentz-boost this particle along the $z$ direction.  The boost matrix is given
in Table~\ref{tab11}, and it takes the form
\begin{equation}\label{boostz}
 B(\eta) = \pmatrix{\exp{(\eta/2)} & 0 \cr 0 & \exp{(-\eta/2)} } .
\end{equation}
Then its momentum becomes
\begin{equation}\label{eta}
  p_{z} = m~\sinh(\eta), \quad\mbox{or}\quad e^{\eta} = \frac{p_{z} + \sqrt{p_{z}^2 + m^2}}{m} .
\end{equation}
This momentum remains invariant under rotations around the $z$ axis.  The rotation
matrix $Z(\phi)$ given in Eq.(\ref{rotz}) commutes with the boost matrix $B(\eta)$ of
Eq.(\ref{boostz}).

The story is different for rotations around an axis perpendicular to the $z$ axis.  Let us
pick the rotation around the $y$ axis given in Eq.(\ref{wm+}).  This matrix can be boosted as
$B(\eta)R(\theta)B^{\dag}(\eta)$, to become
\begin{equation}\label{wm99}
\pmatrix{\cos(\theta/2) & - e^{\eta}\sin(\theta/2) \cr
 e^{-\eta}\sin(\theta/2) & \cos(\theta/2) } ,
\end{equation}
where the boost matrix $B(\eta)$ is that of Eq.(\ref{boostz}). According to Eq.(\ref{eta}),
$\eta$ becomes infinite as the mass becomes smaller.  If we decide to keep all the quantities
in Eq.(\ref{wm99}) finite, the upper-right element $e^{\eta}\sin(\theta/2)$ must be finite.
Let that be $\gamma$.  The lower-left element then becomes
$e^{-2\eta}\gamma$ which vanishes as $\eta$ becomes infinite. The angle $\theta$ becomes zero.
Thus, the boosted rotation matrix becomes the triangular matrix
\begin{equation}
 T(\gamma) = \pmatrix{1 & -\gamma \cr  0 & 1}, \quad\mbox{and}\quad
 \dot{T}(\gamma) = \pmatrix{1 & 0 \cr  \gamma & 1} ,
\end{equation}
which are the triangular Wigner matrices given in Eq.(\ref{wm0}).   When they
are applied to the spinors given in Table~\ref{spinors}, $u$ and $\dot{v}$ remain
invariant, but $\dot{u}$ and $v$ become changed as shown in Table~\ref{trispin}.

\begin{center}
\begin{table}
\caption{$ T(\gamma)$ and $\dot{T}(\gamma)$ transformations on the spinors.
Due to the parity invariance of the Lie algebra of the Lorentz group, we should consider
the triangular matrices and their dots applicable to both $u$ and $v$, and and also to
$\dot{u}$ and $\dot{v}$. }\label{trispin}
\vspace{2mm}
\begin{center}
\begin{tabular}{cccccc}
\hline\\ [-2.3ex]
\hline\\ [-1.2ex]
 \hspace{3mm}  &\hspace{5mm} & $T(\gamma)$ with $+\eta$  & \hspace{5mm}&
 $\dot{T}(\gamma) $ with $-\eta$ \hspace{2mm}\\[1.0ex]
\hline\\[0.6ex]
\hspace{3mm} Spinors  &\hspace{5mm}&
       $ T(\gamma) u = u $ &\hspace{5mm}&  $ \dot{T}(\gamma) u = u + \gamma v $ & \hspace{2mm}
\\ [2.0ex]
 &{}&  $ T(\gamma) v = v - \gamma u  $ &\hspace{5mm}&  $ \dot{T}(\gamma) v = v $ &
 \hspace{2mm}
\\ [2.0ex]
\hline \\[-0.3ex]
\hspace{3mm} Dotted spinors &\hspace{5mm}&
 $T(\gamma)\dot{u} = \dot{u}   $ & \hspace{5mm}   &
     $\dot{T}(\gamma) \dot{u} = \dot{u} + \gamma\dot{v} $ & \hspace{2mm}
     \\[2.0ex]
 &{}& $    T(\gamma)\dot{v} = \dot{v} - \gamma \dot{u} $ & \hspace{5mm}   &
   $\dot{T}(\gamma)\dot{v} = \dot{v}     $ & \hspace{2mm}
     \\[2.0ex]
\hline\\ [-2.3ex]
\hline
\end{tabular}
\end{center}
\end{table}
\end{center}

Here again, there is the rotational degree of freedom around the $z$ axis.  The matrix of Eq.(\ref{wm99})
is generalized into
\begin{equation}\label{wm99z}
\pmatrix{0 & e^{-i\phi/2} \cr e^{i\phi/2} & 0}
 \pmatrix{\cos(\theta/2) & - e^{\eta}\sin(\theta/2) \cr e^{-\eta}\sin(\theta/2) & \cos(\theta/2) }
\pmatrix{0 & e^{i\phi/2} \cr e^{-i\phi/2} & 0},
\end{equation}
which becomes
\begin{equation}\label{wm99z2}
 \pmatrix{\cos(\theta/2) & - e^{-i\phi}e^{\eta}\sin(\theta/2) \cr
  e^{i\phi}e^{-\eta}\sin(\theta/2) & \cos(\theta/2) }  .
\end{equation}
In the large-$\eta$ limit, this expression leads to the triangular matrices of Eq.(\ref{wm0z}).

\section{Representations of Scalars, Vectors, and Tensors for Photon Helicity,
Gauge-dependent Four-Potentials and Gauge-independent Field Tensor}\label{svt}

In the non-relativistic regime the process of constructing three spin-1 states
and one spin-0 state from two spinors is guided by well known spin addition
mechanisms of quantum mechanics~\cite{messiah1962}.
On the other hand in the relativistic regime, due the dotted representation
of $SL(2,c)$ there are two more two-component spinors for each spin-1/2
particle~\cite{knp86,bkn15,berest82,gelfand63,wein64a}.
When all types of the two-component spinors are combined in this regime,
totally there are 16 states.  The construction mechanism of those 16
states are similar to those that are constructed in the $SU(2)$ regime,
and its details were given in our earlier work~\cite{bkn17symm}.
For the purpose of this work it will suffice to give the results in Table~\ref{tab77}.

\begin{table}
\caption{Sixteen combinations of the $SL(2,c)$ spinors. In the non-relativistic sector,
there are two spinors leading to four bilinear forms. In its relativistic extension,
there are two undotted and two dotted spinors. These four-spinors add up
to sixteen independent bilinear combinations~\cite{bkn14symm}.}\label{tab77}
\vspace{3mm}
\begin{center}
\begin{tabular}{ccccccc}
 \hline \\[-2.3ex] \hline\\ [-0.7ex]
{}& Spin 1 & \hspace{10mm} & Spin 0 & {}
\\[0.9ex] \hline\\[-0.9ex]
{}& $ uu,$ \quad $\frac{1}{\sqrt{2}}(uv + vu), $ \quad $ vv, $ & {} & $\frac{1}{\sqrt{2}}(uv - vu) $ & {}
\\[1.7ex] \hline\\[-0.7ex]
{}& $ \dot{u}\dot{u},\quad \frac{1}{\sqrt{2}}(\dot{u}\dot{v} + \dot{v}\dot{u}),
 \quad \dot{v}\dot{v}, $ & {} &
$\frac{1}{\sqrt{2}}(\dot{u}\dot{v} - \dot{v}\dot{u}) $ & {}
\\[1.7ex] \hline\\[-0.7ex]
{}& $ u\dot{u},\quad \frac{1}{\sqrt{2}}(u\dot{v} + v\dot{u}),
\quad v\dot{v}, $ & {} &
$\frac{1}{\sqrt{2}}(u\dot{v} - v\dot{u}) $ & {}
\\[1.7ex] \hline\\[-0.7ex]
{}& $\dot{u}u, \quad \frac{1}{\sqrt{2}}(\dot{u}v + \dot{v}u),
\quad \dot{v}v, $ & {} &
$\frac{1}{\sqrt{2}}( \dot{u}v - \dot{v}u) $ & {}
\\[1.7ex]
\hline\\[-2.3ex]\hline\\[-0.8ex]
\end{tabular}
\end{center}
\end{table}

\par

The spinors in Table~\ref{tab77} can be partitioned into the following states:
\begin{itemize}
	\item[1.] scalar with one state,
	\item[2.] pseudo-scalar with one state,
	\item[3.] four-vector with four states,
	\item[4.] axial vector with four states,
	\item[5.] second-rank tensor with six states.
\end{itemize}

\par
Furthermore, it can be observed from Table~\ref{tab77} that the combinations
\begin{equation}\label{max11}
S = \frac{1}{\sqrt{2}}(uv - vu), \quad\mbox{and}\quad
\dot{S} =  \frac{1}{\sqrt{2}}(\dot{u}\dot{v} - \dot{v}\dot{u})
\end{equation}
are invariant both under rotations and boosts.  Therefore, they are scalars in
the relartivistic world.  Let us next consider the following combinations.
\par
\begin{equation}\label{max12}
S_{+} = \frac{1}{\sqrt{2}}\left(S + \dot{S}\right), \quad\mbox{and}\quad
S_{-} = \frac{1}{\sqrt{2}}\left(S - \dot{S}\right) .
\end{equation}
Under the dot conjugation, $S_{+}$ remains invariant, but $S_{-}$ changes sign.
As was noted in Sec.~\ref{lgroup}, the dot conjugation corresponds to space
inversion.  Therefore, $S_{+}$ is a scalar, while $S_{-}$ is called a pseudo-scalar.

\begin{figure}
\centerline{\includegraphics[scale=4.0]{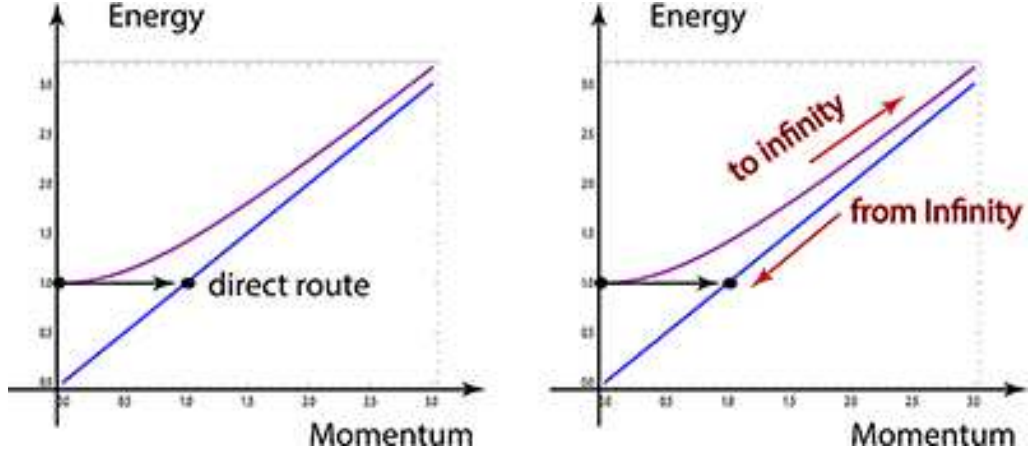}}
\caption{Wigner excursion.  We are interested in transforming a massive particle
at rest into a massless particle with the same energy as illustrated in fig.(a),
but this transformation is not allowed within the framework of the Lorentz group
where the mass is an invariant quantity.  Thus we boost the system to the
infinite-momentum state where the mass hyperbola coincides with the light-cone.
Then we can come back to a finite-momentum state along the light cone as
illustrated in fig.(b)~\cite{bkn14symm}.} \label{excur}
\end{figure}
\subsection{Four-vectors, Four-potential and the Gauge Transformation}\label{4vec}

Let us rewrite the expression for the space-time four-vector given in
Eq.(\ref{slc01}) as
\begin{equation}\label{max21}
\pmatrix{t + z & x - iy \cr x + iy & t - z}.
\end{equation}
It becomes
\begin{equation}\label{max22}
\pmatrix{t - z & -x + iy \cr - x - iy & t + z} ,
\end{equation}
under the parity operation.
The off-diagonal elements undergo sign changes, and the diagonal elements
become interchanged.  We can now construct a four-vector and its
``dot'' conjugation in terms of spinors, in the
following form:
\begin{equation}\label{4vec11}
V \simeq \pmatrix{\dot{v}u - u\dot{v}  & \dot{v}v - v\dot{v} \cr
u\dot{u} - \dot{u}u & v\dot{u} - \dot{u}v} , \qquad
\dot{V} \simeq \pmatrix{v\dot{u} - \dot{u}u  & v\dot{v} - \dot{v}v \cr
\dot{u}u - u\dot{u} &  \dot{v}u - u\dot{v}} ,
\end{equation}
whose transformation properties are
like those of Eq.(\ref{max21}) and Eq.(\ref{max22}), respectively.

Accordingly,  we write the electromagnetic four-potential as
\begin{equation}\label{4vec22}
A = \pmatrix{A_{0} + A_{z} & A_{x} - iA_{y} \cr
                                 A_{x} + iA_{y}& A_{0} - A_{z}}.
\end{equation}
If boosted along the $z$ direction as $B(\eta) A B(\eta)$,  the matrix $A$ becomes
\begin{equation}
A_{\eta} = \pmatrix{ \left(A_{0} + A_{z}\right)e^{\eta} & A_{x} - iA_{y} \cr
  A_{x} + iA_{y}& \left(A_{0} - A_{z}\right) e^{-\eta}} .
\end{equation}
We can then make the Wigner excursion as illustrated in Fig.~\ref{excur}, which
transforms this matrix for a massive particle at rest to that of a massless
particle with the same energy.  The net result is
\begin{equation}\label{4vec33}
A = \pmatrix{A_{0} + A_{z} & A_{x} - iA_{y} \cr A_{x} + iA_{y}& 0},
\end{equation}
resulting in  $A_{0} = A_{z}$ which is widely known as the Lorentz condition.
To the language of spinors this condition is translated as $ v\dot{u} = \dot{u}v$.

If we perform the $T(\gamma)$ on $u$ and $v$, while $\dot{T}(\gamma)$ on
$\dot{u}$ and $\dot{v}$ in $V$ of Eq.(\ref{4vec11}), since $A$ and $V$ have the same
transformation properties, now
the matrix $A$ becomes:
\begin{equation}\label{euc55}
     A + 2\gamma\pmatrix{0 & A_{0}\cr A_{0}   &   0}  .
\end{equation}
This results in the
addition of $2\gamma A_{0} $ to $A_{x}$.  It is a translation in the plane
of $A_{x}$ and $A_{y}$.

On the other hand, if we perform the $\dot{T} (\gamma)$ on $u$ and $v$, \
while $T(\gamma)$ on $\dot{u}$ and $\dot{v}$,  $A$ becomes
\begin{equation}\label{cyl22}
A - 2\gamma \pmatrix{A_{x} & 0 \cr 0 & 0} .
\end{equation}
The triangular matrices $T(\gamma)$ and $\dot{T} (\gamma)$ are given in
Table~\ref{trispin}.

The question next is which one we choose between Eq.(\ref{euc55}) and
Eq.(\ref{cyl22}) for our transformation.    In order to decide, let us go
to the transformation rule given in Eq.(\ref{slc04}) for the four-vector,
and apply the triangular matrix $T(\gamma)$ as the Lorentz transformation
matrix, resulting in the matrix multiplication
\begin{equation}
 \pmatrix{1 & -\gamma \cr 0 & 1}
 \pmatrix{A_{0} + A_{z} & A_{x} - iA_{y} \cr A_{x} + iA_{y}& 0}
 \pmatrix{1 &  0 \cr -\gamma  & 1} ,
  \end{equation}
which becomes
\begin{equation}\label{cyl55}
A - 2\gamma \pmatrix{A_{x} & 0 \cr 0 & 0} .
\end{equation}
This form is the same as the form given in Eq.(\ref{cyl22}).  Thus, we choose
the transformation of Eq.(\ref{cyl22}) as the $T(\gamma)$ and $\dot{T}(\gamma)$
transformations applicable to the spinors.

Both in Eq.(\ref{cyl22}) and Eq.(\ref{cyl55}), $A_{x}$ and $A_{y}$ remain
invariant while there is an additional quantity to $A_{0} + A_{z}$.  The
$T(\gamma)$ matrices therefore leads to a gauge transformation.

What we have done so far can be rotated around $z$ axis.  Then, $\gamma$ is
replaced by $\gamma e^{-i\phi}$.  The transformed $A$ of Eq.(\ref{cyl22}) becomes
\begin{equation}\label{cyl77}
A - 2\gamma \pmatrix{A_{x}\cos\phi + A_{y}\sin\phi & 0 \cr 0 & 0} .
\end{equation}
It is possible to reach the same conclusion using the four-by-four formulation
of the Lorentz group.  This larger representation contains geometries leading to
Eq.(\ref{euc55}) and Eq.(\ref{cyl77})~\cite{kiwi90jmp,kuper76,hk81ajp,wein95}.
We now know from Eq.(\ref{cyl77}) that $T\left(\gamma e^{-i\phi}\right)$ performs
a gauge transformation.

Let us go back to the limiting process discussed in Sec.~\ref{massless}.  According
to Sec.~\ref{massless}, the transverse rotational degrees of freedom collapse into
one gauge degree of freedom in the infinite-momentum or zero-mass limit, as
illustrated in Table~\ref{emc}.  This aspect was observed first by Han {\it et al.}
in 1983~\cite{hks83pl}, and its geometry was given by Kim and Wigner in
1990~\cite{kiwi90jmp}.  The most recent version of this geometry was given by the
present authors in 2017~\cite{bkn17symm}.

The matrices given in Eq.(\ref{4vec11}) were given in our earlier paper on this
subject~\cite{bkn17symm}.  However, as we go deeper into the problem by starting
from the transformation property of each spinor, it was inevitable to make a number
of minus-sign adjustments.  These changes do not alter the conclusions given there
and those given here.

\subsection{Second-Rank Tensor for the Electromagnetic Field and Gauge
Independence}\label{tensor}

As was noted in our earlier publication~\cite{bkn17symm}, it is possible to construct
the 2nd-rank tensor from bilinear combinations of spinors, and the tensor can
take the form
\begin{equation}\label{max59}
\pmatrix{ 0 & -f_z & -f_x & -f_y \cr f_z & 0 & -g_y & g_x \cr
f_x & g_y & 0 & -g_z \cr f_y & -g_x & g_z & 0  } ,
\end{equation}
which can be used for the gauge-invariant electromagnetic field tensor.

Let us first write its $z$ components as
\begin{equation}\label{max51}
f_{z} \simeq \frac{1}{2}\left[(uv + vu) -
 \left(\dot{u}\dot{v} + \dot{v}\dot{u}\right)\right] ,
\qquad
g_{z} \simeq \frac{1}{2i}\left[(uv + vu) +
\left(\dot{u}\dot{v} + \dot{v}\dot{u}\right)\right] .
\end{equation}
These quantities are invariant under the boost along the $z$ direction. They
are also invariant under rotations around this axis, but they are not invariant
under boosts along or rotations around the $x$ or $y$ axis.  The parity
operation on spinors correspond to dot conjugation, and thus $f_{z}$ and $g_{z}$
are respectively anti-symmetric and symmetric under the parity operation,
as in the case of  the electric an magnetic fields.

\par
As to the $x$ and $y$ components, they can be constructed as:
\begin{eqnarray}\label{max55}
&{}& f_{x} \simeq \frac{1}{2}\left[ \left(uu + vv\right) -
 \left(\dot{u}\dot{u} + \dot{v}\dot{v}\right)\right], \nonumber\\[1ex]
&{}& f_{y} \simeq \frac{1}{2i}\left[ \left(uu - vv\right)
- \left(\dot{u}\dot{u} - \dot{v}\dot{v}\right)\right] ,
\end{eqnarray}
and:
\begin{eqnarray}\label{max57}
&{}& g_{x} \simeq \frac{1}{2i}\left[ \left(uu + vv\right) +
 \left(\dot{u}\dot{u} + \dot{v}\dot{v}\right)\right], \nonumber\\[1ex]
&{}& g_{y} \simeq -\frac{1}{2}\left[ \left(uu - vv\right)
+ \left(\dot{u}\dot{u} - \dot{v}\dot{v}\right)\right] .
\end{eqnarray}
At this point, we note that $f_{x}$ and $f_{y}$ are anti-symmetric under dot
conjugation, while $g_{x}$ and $g_{y}$ are symmetric.  The $f_{i}$ of
Eqs.~(\ref{max51}) and (\ref{max55}) transform like a three-dimensional vector,
as in the case of the electric field.  As for $g_{i}$ of Eqs.~(\ref{max51}) and
(\ref{max57}), they remain invariant under the dot conjugation.  They form a
pseudo-vector like the magnetic field.

\par
We can now investigate the symmetry of photons by taking the Wigner excursion
as illustrated in Fig.~\ref{excur}. If, in Eq.(\ref{max55}) and Eq.(\ref{max57}),
we keep only the terms that become larger for larger values of $\eta$, they can
be identified as the transverse components of the electric and magnetic fields
with:
\begin{eqnarray}\label{max60}
&{}& E_{x} \simeq \frac{1}{2} \left(uu - \dot{v}\dot{v}\right),
 \qquad
 E_{y} \simeq \frac{1}{2i} \left(uu + \dot{v}\dot{v}\right),
 \nonumber\\[2ex]
&{}& B_{x} \simeq \frac{1}{2i} \left(uu + \dot{v}\dot{v}\right),
 \qquad
 B_{y} \simeq -\frac{1}{2} \left(uu - \dot{v}\dot{v}\right) .
\end{eqnarray}
The $T(\gamma)$ transformations applicable to $u$ and $\dot{v}$ are the two-by-two
matrices
\begin{equation}\label{3tri09}
\pmatrix{1 & -\gamma \cr 0 & 1}, \quad\mbox{and}\quad
\pmatrix{1 & 0 \cr \gamma & 1} ,
\end{equation}
respectively, as given in Eq.(\ref{wm0}).
Both $u$ and $\dot{v}$ are invariant under these transformations.  Therefore,
these electric and magnetic fields are invariant under the gauge transformation.

The electric and magnetic field components are perpendicular to each other. Furthermore,
\begin{equation}
 B_{x} = E_{y}, \quad\mbox\quad B_{y} = - E_{x} .
\end{equation}
In order to examine how the photon helicity is translated into the electric and magnetic
fields perpendicular to the direction of propagation, let us first construct
\begin{eqnarray}\label{max66}
&{}& E_{+} \simeq E_{x} + iE_{y}, \qquad B_{+} \simeq B_{x} + iB_{y},\\[2ex]
&{}& E_{-} \simeq E_{x} - iE_{y}, \qquad B_{-} \simeq B_{x} - iB_{y}.
\end{eqnarray}
Thus,
\begin{equation}\label{max63}
B_{+} \simeq E_{+} \simeq uu, \qquad B_{-} \simeq E_{-} \simeq \dot{v}\dot{v},
\end{equation}
which means the photon spin is either in the direction of momentum or in the
opposite direction.

Under the parity operation, the direction of momentum is reversed, while $uu$
and $\dot{v}\dot{v}$ become $\dot{u}\dot{u}$ and $vv$ respectively.  This means
that the spinors are replaced by the surviving terms in during the Wigner
excursion in the opposite direction.  The the direction of the spin remains
unchanged.  This is what we expect from the parity rule for momentum and
angular momentum.

Some of the formulas presented in this Subsection are from our previous
publication~\cite{bkn17symm}.  In the present paper, we have given a more
precise definition of the zero-mass limit in terms of the Wigner excursion
as illustrated in Fig.~\ref{excur}, as well as a more detailed application of
the $T(\gamma)$ transformation to each spinor.  In this way, we were able to
study the effect of parity operation on the electromagnetic tensor in terms
of the $SL(2,c)$ spinors.

\subsection{Higher Spins}

Since Wigner's original book of 1931~\cite{wig31,wig59}, the rotation
group, has been extensively discussed in the
literature~\cite{gelfand63,condon51,hamer62}. On the other hand we see that the
Lorentz group has not been fully exploited.  Although there has been some
efforts as to the construction of the most general spin states from the
two-component spinors in the Lorentz-covariant world it has not been examined in
all its details~\cite{berest82}.

All possible spin states that can be constructed from two $SL(2,c)$ spin-1/2 states
are presented in Table~\ref{tab77}.

In the non-relativistic $SU(2)$ regime, the common practice for higher spin constructions
is to resort to the well-known angular momentum addition mechanisms.
For instance, with three spinors it is possible to construct four spin-3/2
states and two spin-1/2 states, which adds up to six states.  Compared to two spin
combinations, this partition process is much more complicated~\cite{fkr71,hkn80ajp}.
In the Lorentz-covariant world, there should be 64 states for three spinors
and 256 states for four spinors.

\begin{figure}
\centerline{\includegraphics[scale=4.0]{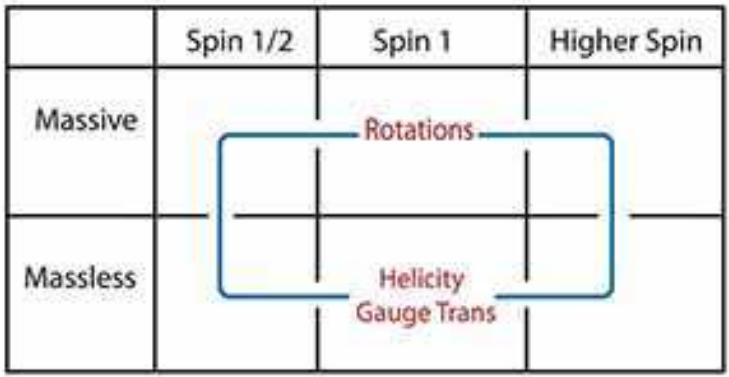}}
\caption{Unified picture of massive and massless particles.
Wigner's little group for massless spin-1 particles generate gauge transformations,
whereas the helicity of the photon is left intact under $T$ and $\dot{T}$.
The handedness of spin-1/2 massless particles is a consequence of
the gauge invariance condition.
}\label{gauge33}
\end{figure}

Since we now know how to Lorentz-boost spinors and take their infinite-$\eta$ limit,
we have a better understanding of the differences between the massive and
massless particle representations and their symmetry properties.
We also observe that the transverse rotations become gauge transformations in the limit
of zero-mass which is basically the infinite-$\eta$ limit. Thus, we are able to combine
them all into the table given in Figure~\ref{gauge33}.

\par
Photons and gravitons are both relativistic integer spin massles particles
that we are focusing on in this work.  As was mentioned in
Subsections~\ref{4vec} and~\ref{tensor}, the observable components have
terms that they become largest for large values of $\eta$.
They are the terms that are invariant under gauge transformations.

\par
Photons have two helicity states. They can be parallel or anti-parallel.
It can be seen from Section~\ref{tensor} that, terms consisting of $uu$
correspond to parallel states while terms with $\dot{v}\dot{v}$ are
for the inti-parallel states.

We have seen in Section~\ref{tensor} that  $uu$ and $\dot{v}\dot{v}$ represent
photon states, whose spins are parallel and anti-parallel to
the momentum, respectively.

\par
In 1964, Weinberg constructed massless particle states~\cite{wein64b},
specifically for photons and gravitons~\cite{wein64c} by
introducing the conditions on the states as:
\begin{equation}\label{weinstates}
N_{1}|\mbox{state}> = 0, \quad\mbox{and}\quad N_{2}|\mbox{state}> = 0,
\end{equation}
where $N_{1}$ and $N_{2}$ are defined in Eq.(\ref{n12}).
Now, we know that $N_{1}$ and $N_{2}$ are the generators of gauge transformations,
therefore the states in Eq.(\ref{weinstates}) are gauge invariant.
Thus, $uu$ and $\dot{v}\dot{v}$ are Weinberg's
states for photons.

Moreover, we can construct $uuuu$ and $\dot{v}\dot{v}\dot{v}\dot{v}$ to represent
spin-2 gravitons.  Since they obey the conditions in Eq.(\ref{weinstates}), they
correspond to Weinberg's graviton states.

\begin{figure}
\centerline{\includegraphics[scale=3.0]{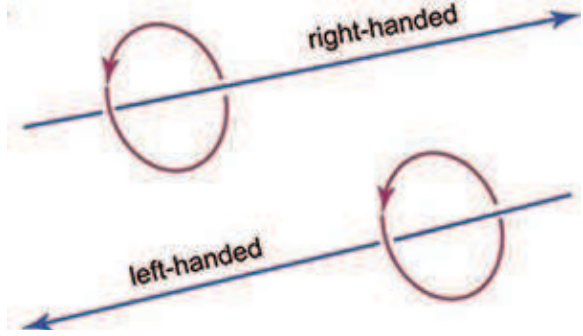}}
\caption{Polarization of massless neutrinos.  This polarization is a consequence
of gauge invariance.}\label{neupol}
\end{figure}

\subsection{Polarization of Massless Neutrinos}\label{neu}
We have established that the triangular matrices $T$ and $\dot{T}$ generate gauge
transformations when applied to four-vectors.  Let us go back to Table~\ref{trispin}.
They also perform gauge transformations on massless spin-half
particles~\cite{hks86jmp,hks82}.

Let us go back to Table~\ref{neupol}.  If we insist on gauge invariance of the world,
massless spin-half particles are polarized.  The dotted particle becomes left-handed,
while the undotted spinor becomes right-handed~\cite{hks86jmp,hks82,png86}.  Indeed,
this is what we observe in the real world. Massless neutrinos and anti-neutrinos are
left- and right-handed respectively.

Yes, neutrinos have non-zero masses~\cite{mohap06,kmn16adv}, but they are so small
compared with their momenta that they can be regarded as small corrections to their
massless states.  In other words, their massless states will play important roles in
physics.

\section*{Concluding Remarks}
From the mathematical point of view, this paper is about the expansion of the group
$SU(2)$ to $SL(2,c)$ within the world of two-by-two matrices. From the physical point
of view, we studied here an issue of building a bridge between Heisenberg's uncertainty
relations and Maxwell's Lorentz-covariant electromagnetic fields.

 It was Einstein who defined the photon as a massless particle in the quantum world
 from his photo-electric effect.  However, he did not consider the ``wings'' of the
 electromagnetic wave.  In the classical picture, there are electric and magnetic
 fields perpendicular to the direction of propagation.  This aspect is translated
 into the polarization of photons.  The question is how?

\par
 This question belongs in the subject area pioneered by Wigner in 1939~\cite{wig39}.
 His 1939 paper deals with the internal space-time symmetries, as specified in
 Table~\ref{emc}.  However, the issue of the electromagnetic four-potential with its
 gauge degree of freedom has a stormy history and was settled in later
 papers~\cite{kiwi90jmp,kuper76,wein95}.  As for the Maxwell tensor, the present
 authors dealt with the problem in their recent publications~\cite{bkn15, bkn17symm}.
 In this paper, we have presented further details of this problem starting from the
 transformation properties of the four spinors defined for the Lorentz-covariant
 world.

\end{document}